\documentclass[authorversion]{sigchi-ext}
\usepackage[T1]{fontenc}
\usepackage{textcomp}
\usepackage[scaled=.92]{helvet} %
\usepackage{graphicx} %
\usepackage{balance}  %
\usepackage{booktabs} %
\usepackage{ccicons}  %
\usepackage{ragged2e} %
\newcommand{\comm}[1]{}

\def\plaintitle{Inventory Systems for Virtual Reality Games} 
\def\emptyauthor{}
\def\plainkeywords{Virtual Reality Games; Inventory; Taxonomy; Interaction; User Interfaces; Virtual Environments; Usability}

\title{Toward a Taxonomy of Inventory Systems for Virtual Reality Games}

\numberofauthors{6}

\author{%
  \alignauthor{%
    \textbf{Sebastian Cmentowski}\\
    \affaddr{University of Duisburg-Essen} \\
    \affaddr{Duisburg, 47057, Germany} \\
    \email{sebastian.cmentowski@uni-due.de} }\alignauthor{%
    \textbf{Andrey Krekhov}\\
    \affaddr{University of Duisburg-Essen}\\
    \affaddr{Duisburg, 47057, Germany}\\
    \email{andrey.krekhov@uni-due.de} } \vfil \alignauthor{%
    \textbf{Ann-Marie M\"uller}\\
    \affaddr{University of Duisburg-Essen}\\
    \affaddr{Duisburg, 47057, Germany}\\
    \email{ann-marie.mueller@stud.uni-due.de} }\alignauthor{%
    \textbf{Jens Kr\"uger}\\
    \affaddr{University of Duisburg-Essen}\\
    \affaddr{Duisburg, 47057, Germany}\\
    \email{jens.krueger@uni-due.de} } \vfil \alignauthor{}\alignauthor{} }

\definecolor{linkColor}{RGB}{6,125,233}
\hypersetup{%
  pdftitle={\plaintitle},
  pdfauthor={\emptyauthor},
  pdfkeywords={\plainkeywords},
  bookmarksnumbered,
  pdfstartview={FitH},
  colorlinks,
  citecolor=black,
  filecolor=black,
  linkcolor=black,
  urlcolor=linkColor,
  breaklinks=true,
}

\begin{document}

\CopyrightYear{2019}
\conferenceinfo{CHI PLAY EA '19}{October 22--25, 2019, Barcelona, Spain}
\setcopyright{rightsretained}
\isbn{978-1-4503-6871-1/19/10}
\doi{https://doi.org/10.1145/3341215.3356285}
\copyrightinfo{\acmcopyright}

\maketitle

\RaggedRight{} 

\begin{abstract}
Virtual reality (VR) games are gradually becoming more elaborated and feature-rich, but fail to reach the complexity of traditional digital games. One common feature that is used to extend and organize complex gameplay is the in-game inventory, which allows players to obtain and carry new tools and items throughout their journey. However, VR imposes additional requirements and challenges that impede the implementation of this important feature and hinder games to unleash their full potential. Our current work focuses on the design space of inventories in VR games. We introduce this sparsely researched topic by constructing a first taxonomy of the underlying design considerations and building blocks. Furthermore, we present three different inventories that were designed using our taxonomy and evaluate them in an early qualitative study. The results underline the importance of our research and reveal promising insights that show the huge potential for VR games.
\end{abstract}





\section{Motivation}
More and more players immerse themselves in virtual environments. Head-mounted displays (HMDs) and tracked hand controllers enable unmatched levels of agency and interactivity and can provide an enjoying experience in virtual reality (VR). In an effort to enrich the player experience even further, latest research work has already tackled a broad set of challenges: Novel navigation techniques~\cite{cmentowski2019outstanding, krekhov2018gullivr} allow the players to explore vast open worlds on their own and intuitive controller designs~\cite{krekhov2017self, zenner2019drag} transform these worlds into interesting and interactive experiences.\par

\begin{marginfigure}[-12.75em]
  \begin{minipage}{0.95\marginparwidth}
    \centering
    \includegraphics[width=\marginparwidth]{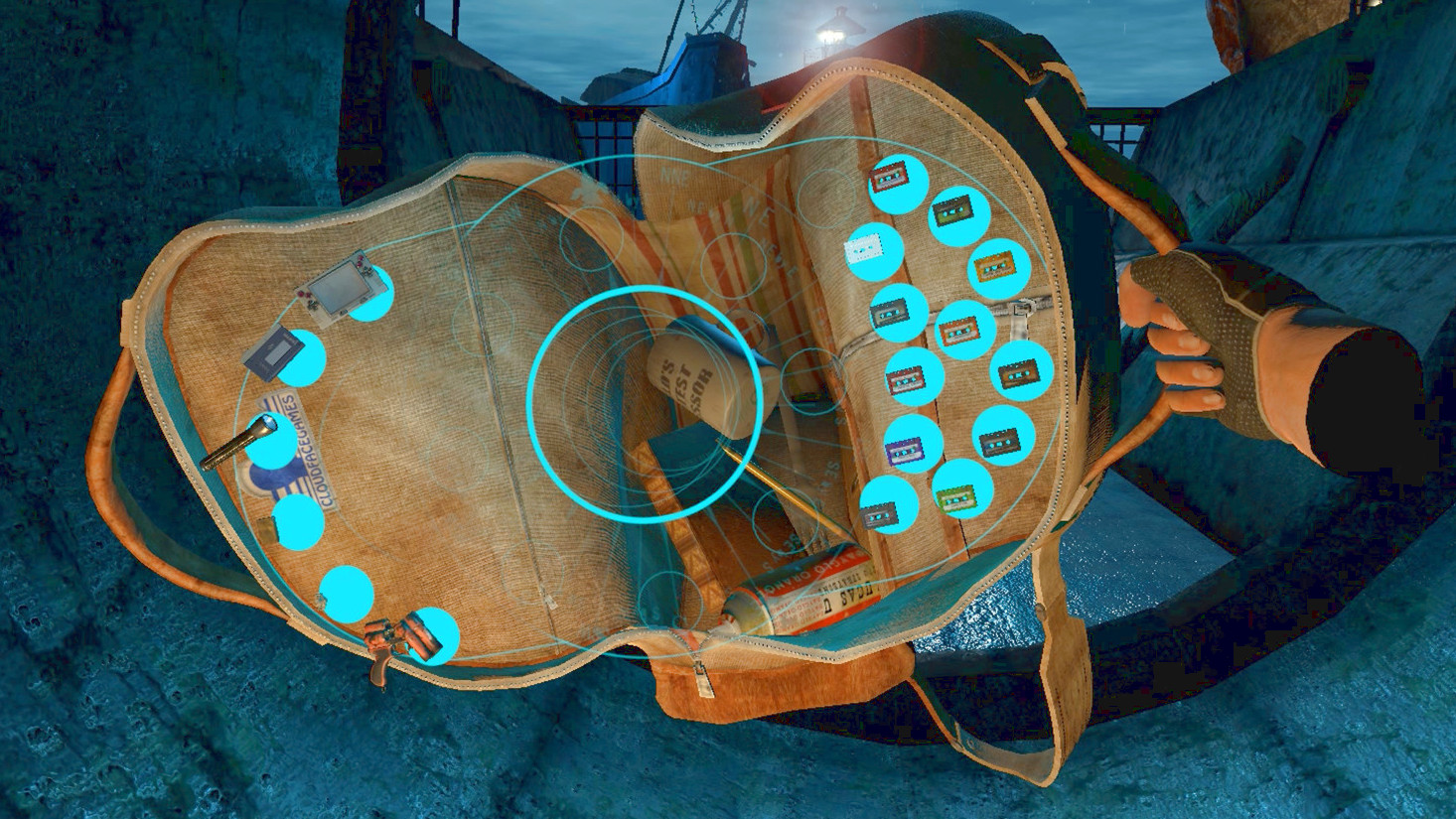}
    \caption[Virtual backpack used as inventory in the game \textit{The Gallery}]{Virtual backpack used as inventory in the game \textit{The Gallery - Episode 1: Call of the Starseed}~\cite{GameGallery}.}~\label{fig:screenshotGallery}
  \end{minipage}
\end{marginfigure}

\begin{marginfigure}[10pt]
  \begin{minipage}{0.95\marginparwidth}
    \centering
    \includegraphics[width=\marginparwidth]{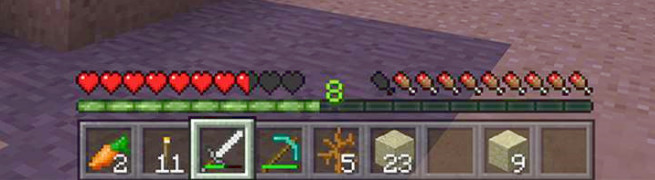}
    \caption[The linear quick bar for frequently used items in the game \textit{Minecraft}]{The linear quick bar for frequently used items in the game \textit{Minecraft}~\cite{GameMinecraft}.}~\label{fig:screenshotMinecraft}
  \end{minipage}
\end{marginfigure}

\begin{marginfigure}[10pt]
  \begin{minipage}{0.95\marginparwidth}
    \centering
    \includegraphics[width=\marginparwidth]{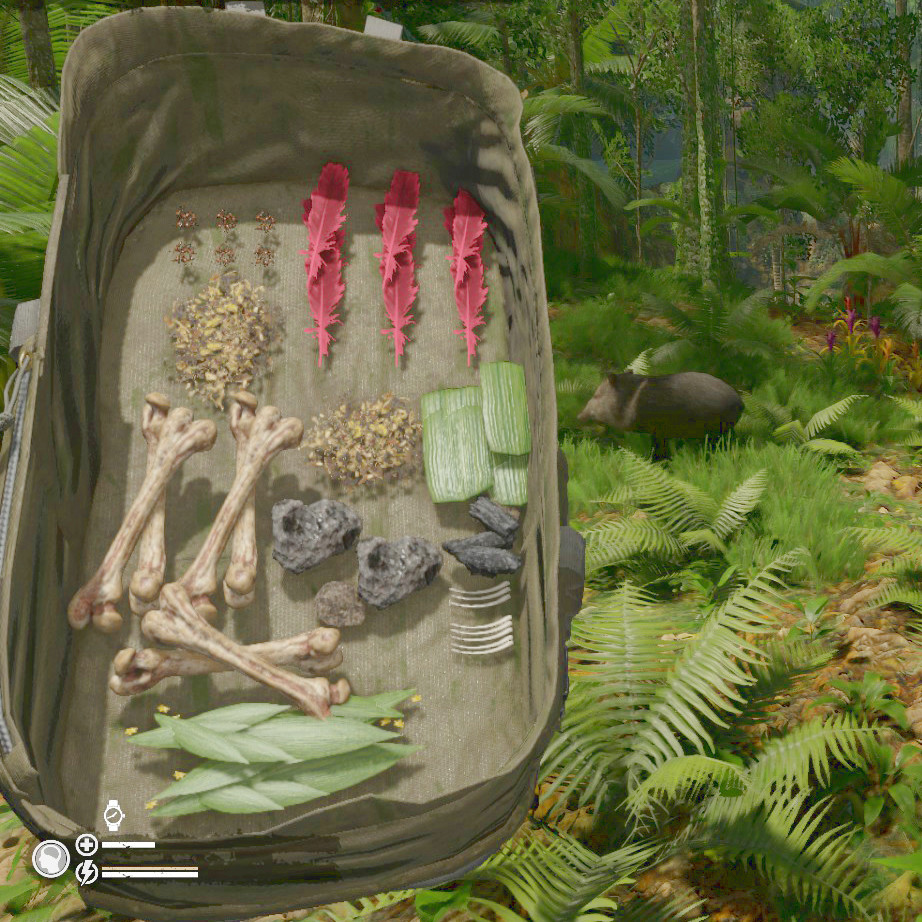}
    \caption[The inventory of the game \textit{Green Hell} is structured by the player alone.]{The inventory of the game \textit{Green Hell}~\cite{GameGreenHell} is structured by the player alone.}~\label{fig:screenshotGreenHell}
  \end{minipage}
\end{marginfigure}

However, many VR games still fail to reach the complexity of traditional digital games. One example of typically overlooked features is the in-game inventory~\cite{wegner2017comparison}. It allows players to acquire new tools and goods throughout their journey and plays an important role in character development. In this work, we establish a starting point for this important topic. Our main contribution is the construction of a first taxonomy covering the broad design space of inventories in VR games, including the different requirements and design possibilities. Based on this preliminary taxonomy, we present three different inventory systems: \textit{Flat Grid}, \textit{Virtual Drawers}, and \textit{Magnetic Surface}. The prototypes were evaluated in a small-scale qualitative study to gather first insights. The player feedback is used to generate early design implications and potential future research questions.

\section{Design Considerations for Inventories}
The use of inventories dates back to the beginning of digital games as such~\cite{GameHamurabi, GameOregonTrail}. They have yet evolved into a standard feature that is used in a majority of current games. Even though the use of VR introduces additional challenges and novel design possibilities, the major purpose of these interfaces stays the same: storing items. The success of a particular implementation depends on various design decisions that impact four crucial factors: \textit{comprehensibility}, \textit{interactivity}, \textit{contextual embedding}, and \textit{personalization}.\par
\textbf{Comprehensibility.} The information for every item that is currently stored in an inventory needs to be displayed in a comprehensible manner. Displaying too many items or meta-information at once on a limited screen can easily lead to visual cluttering. This increases the necessary mental effort and processing time exponentially and can easily spoil the whole gameplay~\cite{Sztajer2010Inventory}.\par
\textbf{Interactivity.} In general, managing and interacting with the inventory and the stored items should be as easy and quick as possible. In extreme cases, situation-dependent controls could be used to simplify the necessary user actions to a minimum. However, such game designs bear the risk of reducing the feeling of agency~\cite{Sztajer2010Inventory}. Especially, in VR games, players like to interact with the environment and to feel in control over the resulting actions. \par
\textbf{Contextual Embedding.} Ultimately, each inventory needs to fit the game's context and purpose. Most implementations can be subdivided into two general groups: The \textit{carry} mode focuses on few, quickly accessible items, whereas the \textit{loot} mode requires easily manageable storage~\cite{hamari2010game, Sztajer2010Inventory}. A typical example of a pure \textit{carry} mode is the game \textit{Fortnite}~\cite{GameFortnite}, whereas the inventory system of \textit{Diablo III}~\cite{GameDiablo3} is a perfect remedy for \textit{loot}-based games.\par
\textbf{Personalization.} The role of inventories as personal and individual space has a huge impact on the game experience. Especially in linear games, players usually have little to no chance to individualize their gameplay. The only true exception is an inventory, where they have complete freedom over content and organization~\cite{Sztajer2010Inventory}. Therefore, freely manageable inventories can provide significant advantages to character identification and presence.

\section{Building Blocks of Inventories for VR}

\marginpar{%
  \fbox{%
    \begin{minipage}{0.95\marginparwidth}
        
      \textbf{Taxonomy of Inventories} \\
      \vspace{1pc} 
      \textbf{(1) Inventory Interface:} 
      \begin{itemize}[leftmargin=*] \compresslist
          \item \textit{type:} overlays, virtual objects, real proxies
          \item \textit{position:} static or moveable
          \item \textit{style:} thematic or realistic
      \end{itemize}
      
      \vspace{0.25pc} \textbf{(2) Item Representation:}
        \begin{itemize}[leftmargin=*] \compresslist
          \item \textit{design:} realistic or abstract
          \item \textit{scale:} scale-preserving, miniaturizing, normalizing
          
      \end{itemize}
      \vspace{0.25pc} \textbf{(3) Item Arrangement:}  \begin{itemize}[leftmargin=*] \compresslist
          \item \textit{layout:} linear, grid, ring, slots, unrestricted
          \item \textit{capacity:} unlimited, fixed, dynamic
          \item \textit{ordering:} unstructured, sortable, sorted
          \item \textit{improvements:} categories, hierarchies, stackability
      \end{itemize}
       \vspace{0.25pc} \textbf{(4) Interaction:} 
       \begin{itemize}[leftmargin=*] \compresslist
          \item \textit{open/close:} click or gesture
          \item \textit{adding items:} automatic or manual
          \item \textit{item manipulation:} raycast-based or physical action
      \end{itemize}
      \includegraphics[width=0.95\marginparwidth]{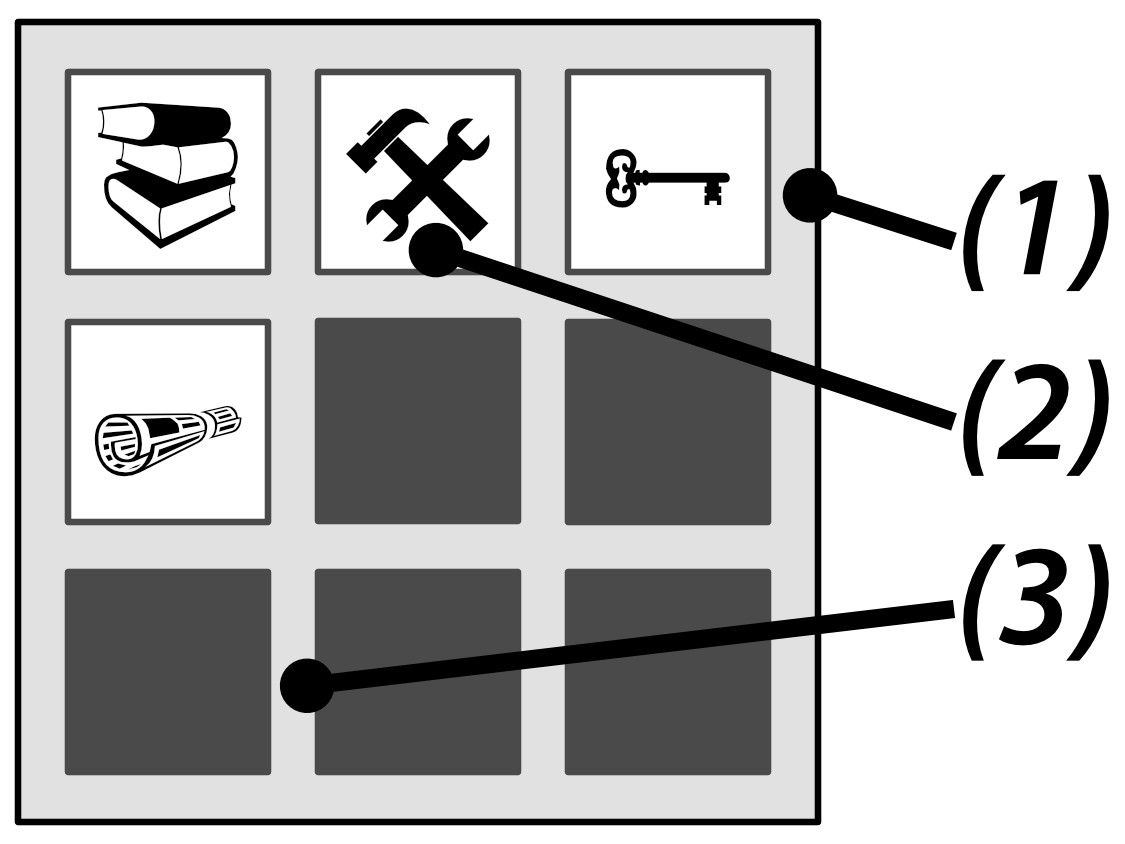}
    \end{minipage}}\label{sec:sidebar} }
    
Inventories are more than simple user interfaces and consist of multiple different components that need to be taken into consideration. In this section, we decompose the inventory into atomic building blocks for a better understanding and analysis.

\subsection{Inventory Interface}
The interface is the basic building block that contains all storage items and determines the position, shape, and design of the whole inventory. In contrast to traditional digital games, the use of VR introduces a whole range of possible interface types. Apart from simple overlays such as the 2D user interface in \textit{SteamVR Home}~\cite{steamVR}, the inventory could be presented as a virtual object in the world. This includes moveable items such as the backpack inventory in \textit{The Gallery}~\cite{GameGallery} (cf. Figure~\ref{fig:screenshotGallery}) and fixed access points, e.g, chests or mailboxes. Alternatively, it is possible to attach the inventory directly to the player. A typical example is the virtual belt in the game \textit{Batman: Arkham VR}~\cite{GameArkhamVR}. Another intriguing approach to further increase realism and agency is the use of a collection of real physical proxies~\cite{krekhov2017self, zenner2019drag}.\par

Inventories in VR games must be placed in a 3D virtual environment which introduces the depth as an additional parameter. Depending on the positioning in relation to the player, the inventory might be occluded by the surrounding or could obstruct the player's view. Both cases usually lead to frustration and decreased usability. Therefore, it might be favorable to give the player the chance to move the storage freely or hide it completely. This feature could also provide further benefits in terms of personalization and interactivity.\par

Usually, VR applications would try to maximize the player's immersion into the virtual scenery and to avoid any thematic cuts. Therefore, it seems natural to match the inventory's style as closely as possible to the surrounding environment. On the other hand, abstract menus offer the advantage of prior knowledge: Most people have already experienced a storage interface of any kind and should be proficient to a certain degree. Therefore, a more abstract design could help to reduce the necessary cognitive load.

\subsubsection{Item Representation}
Every in-game item placed within the inventory needs a graphical or textual representation. The chosen concept strongly influences the type and amount of information that is conveyed to the player. Realistic designs allow for detailed conclusions on the object's shape and physical properties, while also merging into the virtual scenario. In return, a more abstract style, e.g., icons or texts, reduce the visual clutter and make it possible to increase the information density for each item. It is to note that the choice between a 2D and 3D representation is not directly linked to the degree of realism. Instead, this decision is usually based on the type of inventory interface being used.\par

One key aspect of the chosen representation is the displayed size: Preserving the original physical scale of objects does not only allow players to draw conclusions on relative ratios and actual sizes but makes it possible to remove any discrepancy between the real item and its representation in the inventory. However, this design choice bears the risk to quickly fill the limited inventory space with large items and introduce new problems such as item occlusion. The alternative is to miniaturize or normalize the item's scale when adding the object to the inventory.

\subsubsection{Item Arrangement}
The arrangement of the various items in the inventory is at least as important as the representation itself. Depending on the actual use case, the layout should be focused on either accessibility, management, or overview of the content. In digital games, the most common choices are linear shapes for \textit{carry}-inventories, such as the quick bar in \textit{Minecraft}~\cite{GameMinecraft} (cf. Figure~\ref{fig:screenshotMinecraft}), and grids, e.g., in \textit{World of Warcraft}~\cite{gameWoW}, when dealing with larger amounts of items. Other possible designs include ring menus or fixed slots. For immersive experiences aiming to increase presence and personalization, it seems likely to provide more freedom to the player. However, free object placement bares the risk to produce obstructive and chaotic inventories that miss the forced ordering through grids. Nevertheless, this drawback could be used as a gameplay mechanic, like in the survival game \textit{Green Hell}~\cite{GameGreenHell} (cf. Figure~\ref{fig:screenshotGreenHell}).\par

\begin{marginfigure}[-12.15em]
  \begin{minipage}{0.95\marginparwidth}
    \centering
    \includegraphics[width=\marginparwidth]{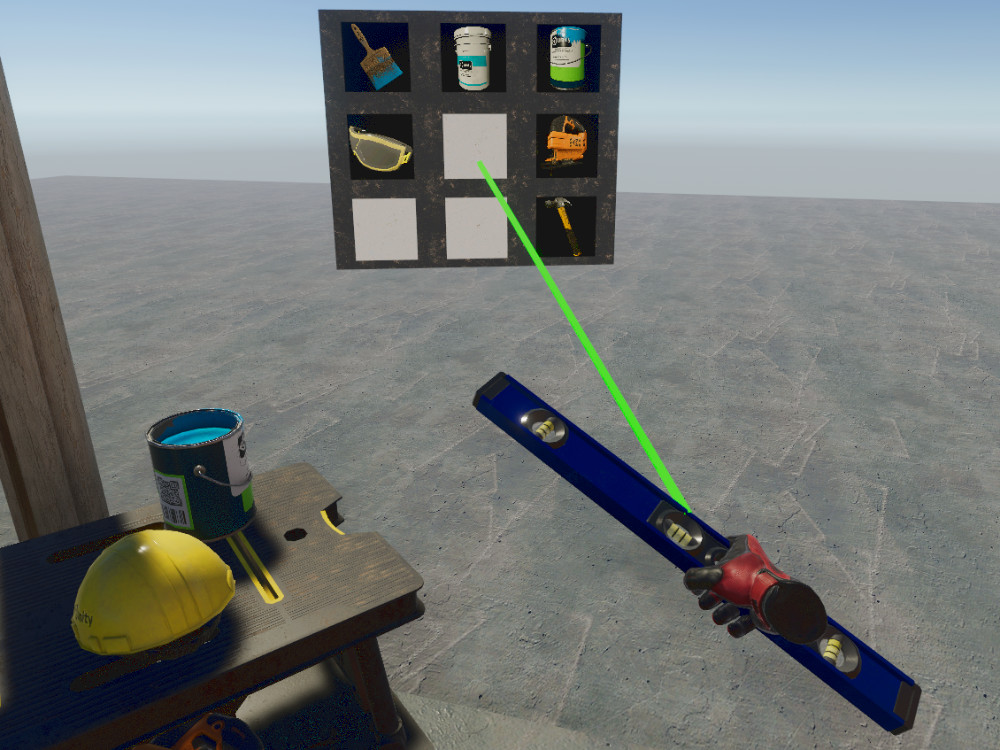}
    \caption{Items are added to the \textit{Flat Grid} using raycast aiming.}~\label{fig:flat1}
  \end{minipage}
\end{marginfigure}

\begin{marginfigure}[10pt]
  \begin{minipage}{0.95\marginparwidth}
    \centering
    \includegraphics[width=\marginparwidth]{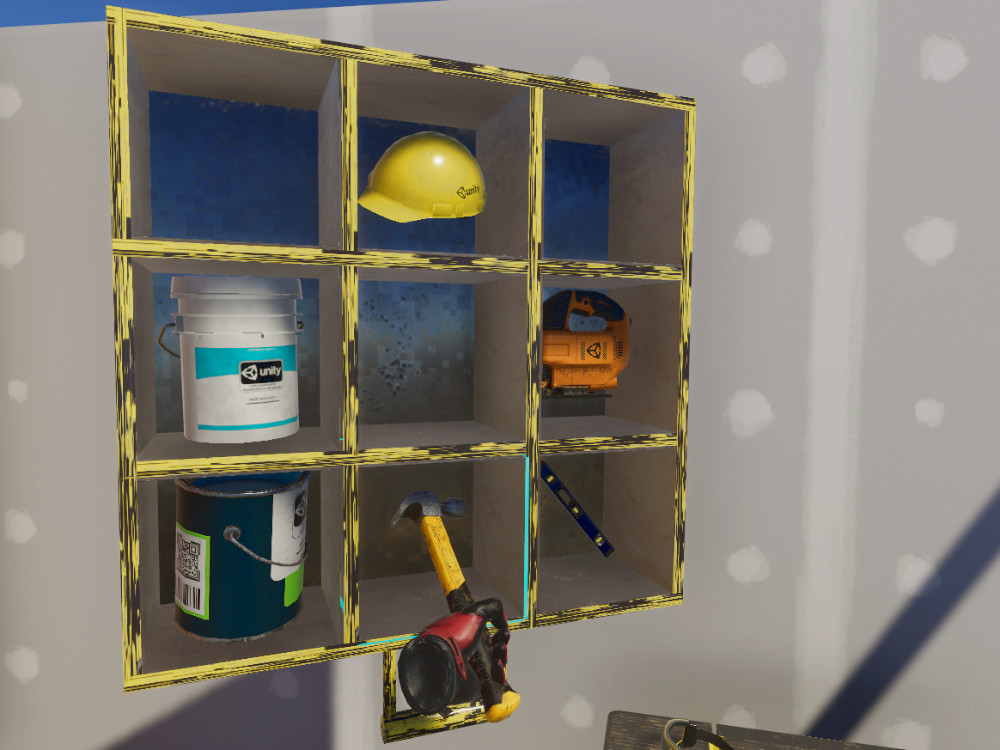}
    \caption{The \textit{Virtual Drawers} contain equally-scaled items.}~\label{fig:drawers1}
  \end{minipage}
\end{marginfigure}

\begin{marginfigure}[10pt]
  \begin{minipage}{0.95\marginparwidth}
    \centering
    \includegraphics[width=\marginparwidth]{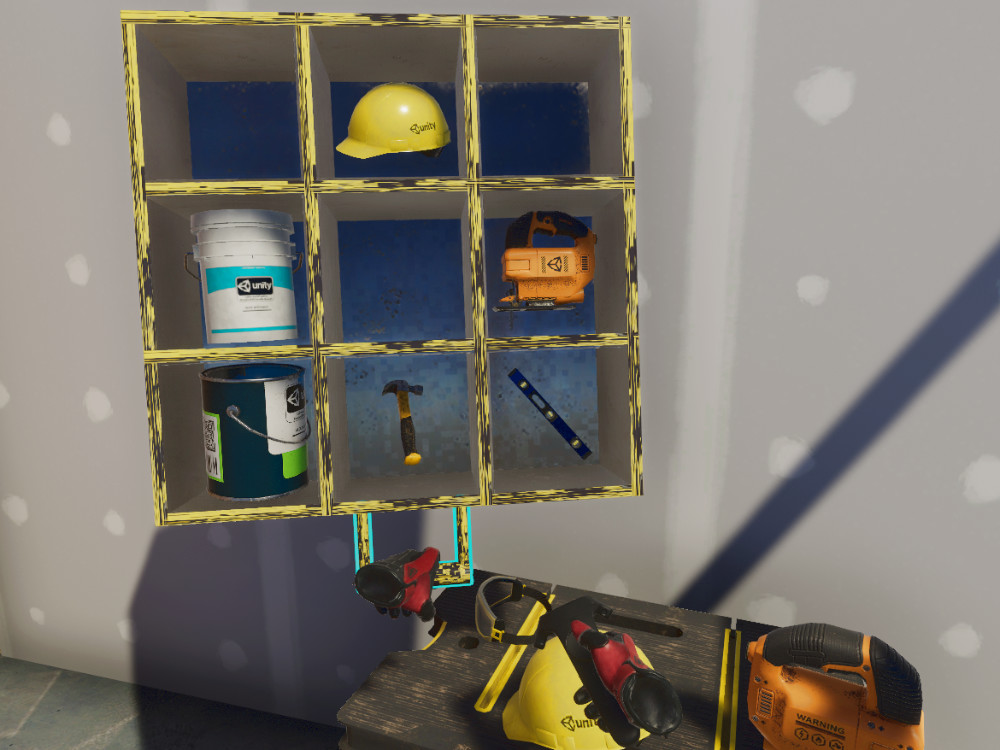}
    \caption{Players can move the inventory using a handle.}~\label{fig:drawers2}
  \end{minipage}
\end{marginfigure}

A common problem of many games is a cluttered inventory providing poor comprehensibility and overview. Reasons are either massive storage sizes or extreme information densities. The easiest solution is a limitation of the inventory capacity or a reduction of the provided information per item. Many games decide to include an additional feature to extend the maximum storage during the journey. However, this step should be combined with an increase of the interface size to preserve the relative information density.\par
In addition to the particular item arrangement strategy, inventories provide different amounts of automatic ordering. This ranges from complete flexibility and self-administered organization, through optional sorting techniques, to fixed arrangements such as the LIFO (last in, first out) approach. After all, inventories in VR should maximize the player's freedom and control over the item arrangement without sacrificing too much comprehensibility. The overview could be improved by including optional concepts such as different categories, hierarchies, or stackability of items. Nonetheless, only the minority of these features have been used in VR games yet and it remains to be investigated whether these can be combined with other aspects such as realistic item representation or free object arrangement.

\subsubsection{Interaction with Inventory}
VR offers completely new interaction approaches using tracked controllers and spatial movements. Therefore, the underlying interactions with the inventory are by far the most intriguing difference in comparison to non-VR games. Instead of using a simple button click, games could implement more natural and interactive ways to open the inventory. A perfect example is \textit{The Gallery}~\cite{GameGallery}, where players reach behind themselves to retrieve a virtual backpack. Additionally, it is easy to give the players full control over the position and orientation of the inventory and allow them to drag it around freely. This opens novel design concepts and provides a very high level of interactivity.\par
The by far most important mechanism is adding and removing items in the inventory. Many non-VR games tend to collect items automatically. However, this approach is less favorable for VR-games. Instead, they could greatly profit from a manual pick-up-process to improve the player's agency. This could be achieved either through more abstract raycast-aiming or more natural and physical gestures. Especially direct interactions with virtual objects outperform distant object selection and manipulation significantly~\cite{mine1997moving}.

\section{Inventory Designs}
The main goal for designing different inventories was to cover the most interesting aspects of the taxonomy and gather first valuable insights. The chosen number of three implementations avoids binary opinions and instead encourages players to provide detailed and in-depth feedback. At the same time, the limited amount prevents subjects from getting lost and potentially frustrated. Our designs differ in various aspects, such as the interface type, item style, or underlying interaction concept. To avoid arbitrary feedback and ensure comparable results, we defined a set of three rules that are shared between all prototypes: 
\begin{enumerate} \compresslist
    \item A realistic and consistent style is preserved.
    \item The inventory is activated with a single button click and starts at a fixed position in the virtual world depending on the player's view.
    \item There are no additional features such as sorting, categories, hierarchies, or item stackability.
\end{enumerate} 

\begin{marginfigure}[-1.6em]
  \begin{minipage}{0.95\marginparwidth}
    \centering
    \includegraphics[width=\marginparwidth]{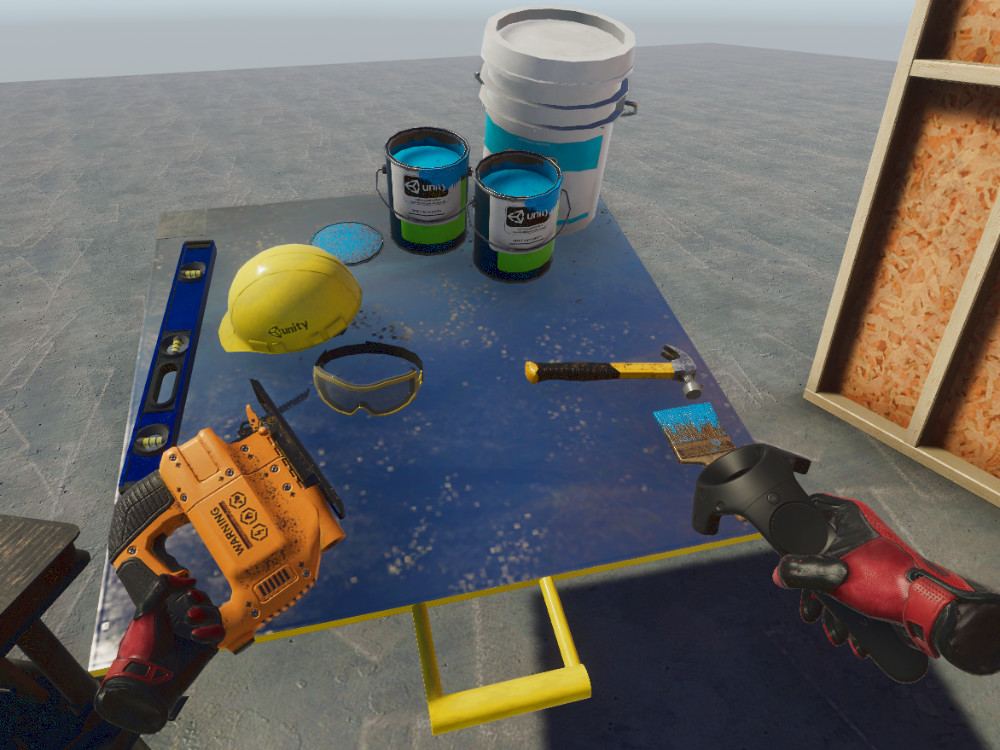}
    \caption{The \textit{Magnetic Surface} allows for fine-graded positioning.}~\label{fig:magnet1}
  \end{minipage}
\end{marginfigure}

\begin{marginfigure}[10pt]
  \begin{minipage}{0.95\marginparwidth}
    \centering
    \includegraphics[width=\marginparwidth]{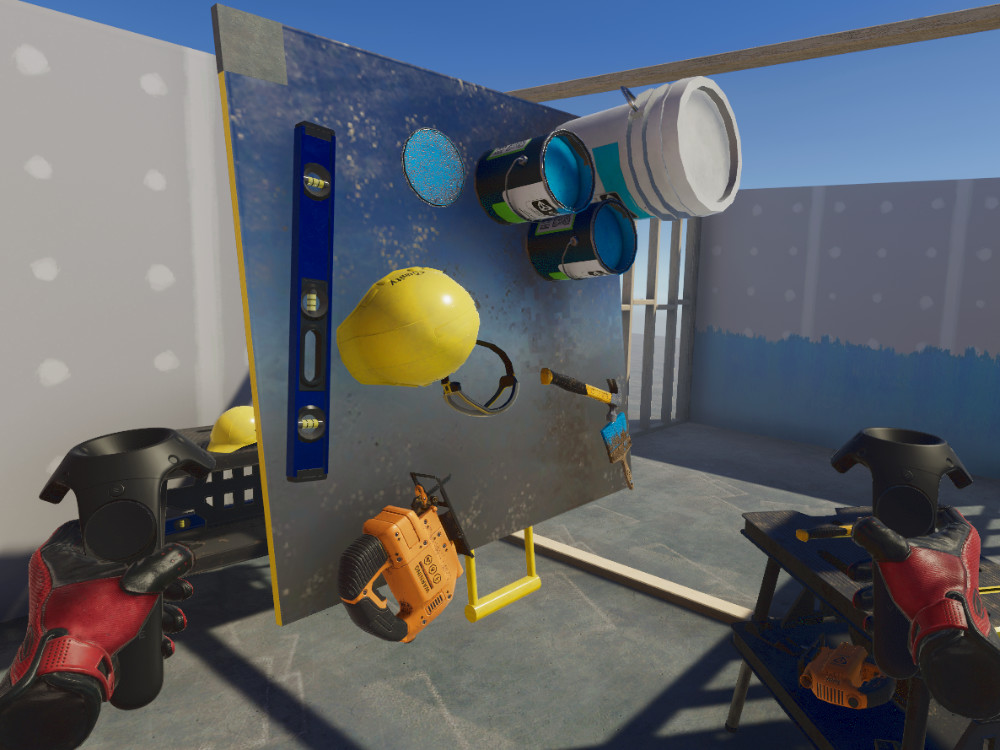}
    \caption{Items stick to the \textit{Magnetic Surface} until removed.}~\label{fig:magnet2}
  \end{minipage}
\end{marginfigure}

\begin{marginfigure}[10pt]
  \begin{minipage}{0.95\marginparwidth}
    \centering
    \includegraphics[width=\marginparwidth]{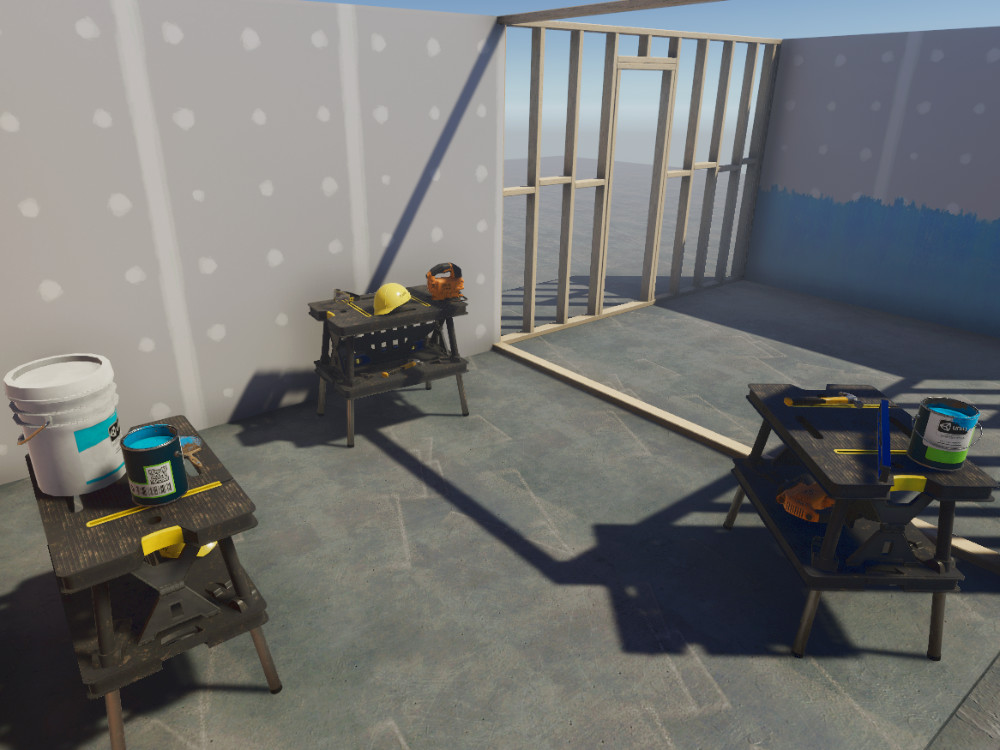}
    \caption{The construction scenario used as testbed game.}~\label{fig:testbed}
  \end{minipage}
\end{marginfigure}

\subsection{Flat Grid}
The first inventory design is a flat two-dimensional overlay that is placed at a static position $1.5m$ in front of the player's head upon activation. The stored items are arranged using a regular grid of three times three virtual slots. This fixed size is chosen to provide some degree of variety for object placement while also mimicking the case of a limited inventory. Each object in the inventory is presented as a realistic 2D image. This prototype closely matches the most common menus in current VR games (cf. \textit{SteamVR Home}~\cite{steamVR}). It should provide a simple and comprehensible design that lacks realism and free object placement.\par

The interaction with the inventory uses raycast pointing: Players could point at an occupied slot and press the button for grabbing objects to retrieve the item. Similarly, releasing an object while pointing at an empty slot would insert the item into the inventory (cf. Figure~\ref{fig:flat1}). This interaction design is consistent with the other prototypes in terms of the necessary buttons and expected behavior but replaces natural grabbing through an abstract raycast aiming.

\subsection{Virtual Drawers}
The second design shares the same grid of nine slots as the \textit{Flat Grid} inventory. However, it is displayed as a virtual 3D shelf floating in the world (cf. Figure~\ref{fig:drawers1}). The stored items are placed on one of the shelf boards as realistic 3D shapes that are scaled to a normalized size to fit the available space. The items are added to the inventory by placing them physically into one of the slots instead of using raycasts. Players have the opportunity to grab the whole inventory using a handle located at one side and move it freely within the environment (cf. Figure~\ref{fig:drawers2}). In contrast to the \textit{Flat Grid}, this design features a more realistic appearance and includes more natural and direct interaction.

\subsection{Magnetic Surface}
The last inventory is a simple metallic work-plate floating at a convenient position to the player's side (cf. Figure~\ref{fig:magnet1}). Similarly to the \textit{Virtual Drawers}, the plate can be positioned freely in the world using an attached handle. This inventory provides the most freedom and control to the user: it avoids any forced organization in the form of grids but uses a magnetic force to stick all stored items to the surface of the work-plate (cf. Figure~\ref{fig:magnet2}). Players can simply put any object onto the plate using natural interaction and it will preserve the chosen position and rotation, as well as the item's original scale and shape. Consequently, this inventory has no maximum capacity or fixed density. Instead, the players are in full control over item arrangement and positioning. They can even use the physical properties of held items to push other stored items around. 

\section{Evaluation and Discussion}
\begin{marginfigure}[0.25em]
    \begin{minipage}{\marginparwidth}
    \centering
    \includegraphics[width=\marginparwidth]{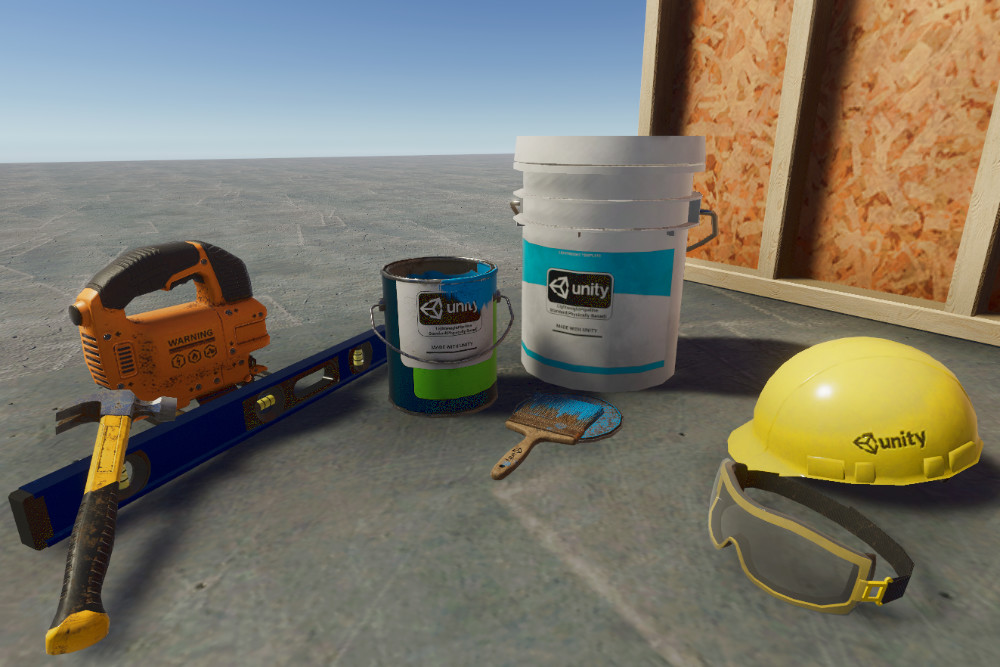}
    \caption{Players used various items to test the inventories.}~\label{fig:items}
  \end{minipage}
\end{marginfigure}

We executed an early qualitative study with 8 participants (3 female, mean age: $29.1$, $SD = 6.19$) to gather insights into how players would use and experience the different inventories. The used scenario was a virtual construction ground featuring all kinds of tools and materials (cf. Figure~\ref{fig:testbed}). This provided a suitable testbed for the evaluation as it allowed the participants to interact with various different items and simulate different uses of inventories (cf. Figure~\ref{fig:items}). During the study, the subjects were briefly introduced to all three prototypes in random order. They were encouraged to explore the different features on their own and provide feedback in an audio-recorded think-aloud process. After examining all aspects of the current inventory to a full degree, the subjects could move on to the next implementation while having the option to return at a later time.\par

\marginpar{\vspace{2em}%
    \begin{minipage}{0.9\marginparwidth}
\textit{"The abstract nature of the flat grid somehow contradicts my expectations towards an immersive VR-game."~(\textbf{P1})}

\vspace{4em}
\textit{"I would love to empty the virtual drawers by turning them upside down."~(\textbf{P2})}

\vspace{4em}
\textit{"I miss the stackability of similar items."~(\textbf{P5})}

\vspace{4em}
\textit{"The magnetic surface is like a scratchpad where I can arrange everything to my needs."~(\textbf{P8})}
\vfill

    \end{minipage}~\label{sec:sidebar2} }

In general, the feedback for all inventories was very positive. Every participant was able to learn the underlying concepts quickly and provide detailed feedback. One major concern with all designs was the occlusion problem between environment and inventory. As VR inventories have to be placed within a 3D scenario, they can easily obstruct important content or appear behind other objects. This weakness was partially fixed by the option to grab and move the inventory to custom spots. This technique allowed for more personalization and was the most appreciated feature throughout the study. An interesting request from one participant was to reduce the setup time by \textit{"restoring the previously adjusted position relative to the head"(\textbf{P3})}.\par

Being asked for their personal favorites, almost all participants gave a similar ranking beginning with the \textit{Flat Grid} as the least interesting choice and the \textit{Magnetic Surface} as the clear winner. At first sight, this contradicts the feedback for the \textit{Magnetic Surface} being described as the least scaleable, structured, and productive of all three inventories. However, players clearly emphasized the importance of interactivity, authenticity, and realism. One of the explicitly-named influencing factors was the shape- and size-preserving behavior of the \textit{Magnetic Surface} prototype. A common suggestion was to implement this feature into the \textit{Virtual Drawers} by \textit{"resizing the shelf boards automatically to fit any item at its original scale"(\textbf{P5})}.\par

Even though most VR applications use raycasts to interact with menus, most participants initially tried to use the \textit{Flat Grid} by placing items with physical grabbing. This is a clear sign towards the superiority of natural interactions.\par

Finally, we asked the subjects to propose use-cases for every inventory. The broad consensus was to choose the \textit{Flat Grid} for efficiency in larger and more loot-based games. One participant mentioned an interesting approach to increase the performance even further: \textit{"inserting an item to an occupied slot should swap both items to avoid additional steps"(\textbf{P1})}. The virtual drawers were generally preferred as a more interactive alternative that is best suited for larger worlds and manageable item counts. Both approaches were described as \textit{"storage inventories"(\textbf{P7})} for less often used items. Finally, the \textit{Magnetic Surface} was sometimes called \textit{"working storage"(\textbf{P8})} and was preferred for few regularly used carry-items. Interestingly, some participants liked the idea of using this inventory in multiplayer games to provide mutual storage shared by all players.

\section{Conclusion and Future Work}
We proposed a first step toward the use of inventories in virtual environments. Our preliminary taxonomy explains the necessary considerations, building blocks, and potential design choices to be considered when creating inventories for VR games. Our qualitative study supports our basic assumption that these interfaces could provide novel game mechanics and improve game experience and enjoyment. The detailed player feedback revealed first interesting insights backing our taxonomy approach and hinting toward additional parameters and design considerations such as the occlusion problem. In our future work, we will refine and extend the existing taxonomy to incorporate additional player and developer input. Additionally, we suggest investigating the overall significance of storage systems in VR games to explore the effects of these interfaces on the game experience. The ultimate goal is to build a fundamental research body on VR inventories incorporating a comprehensive taxonomy and a set of design guidelines to be used by researchers and practitioners.

\comm{We proposed a first step toward the use of inventories in virtual environments. Our preliminary taxonomy explains the necessary considerations, building blocks, and potential design choices to be considered when creating inventories for VR games. Our qualitative study supports our basic assumption that these interfaces could provide novel game mechanics and improve game experience and enjoyment. The detailed player feedback revealed first interesting insights toward future design guidelines. As a future step, we suggest investigating the overall significance of storage systems in VR games to explore the effects of these interfaces on the game experience. The ultimate goal is to generate a comprehensive set of design guidelines to be used by researchers and practitioners.}

\balance{} 

\bibliographystyle{SIGCHI-Reference-Format}
\bibliography{literature}


\begin{thebibliography}{00}

%%% the Web address.

\ifx \showCODEN    \undefined \def \showCODEN     #1{\unskip}     \fi
\ifx \showDOI      \undefined \def \showDOI       #1{{\tt DOI:}\penalty0{#1}\ }
  \fi
\ifx \showISBNx    \undefined \def \showISBNx     #1{\unskip}     \fi
\ifx \showISBNxiii \undefined \def \showISBNxiii  #1{\unskip}     \fi
\ifx \showISSN     \undefined \def \showISSN      #1{\unskip}     \fi
\ifx \showLCCN     \undefined \def \showLCCN      #1{\unskip}     \fi
\ifx \shownote     \undefined \def \shownote      #1{#1}          \fi
\ifx \showarticletitle \undefined \def \showarticletitle #1{#1}   \fi
\ifx \showURL      \undefined \def \showURL       #1{#1}          \fi

\bibitem{gameWoW}
{{Blizzard Entertainment}}. 2004.
\newblock \emph{World of Warcraft}.
\newblock Game [PC].   (23 November 2004).
\newblock
\newblock
\shownote{Vivendi, Paris, France.}


\bibitem{GameDiablo3}
{{Blizzard Entertainment}}. 2012.
\newblock \emph{Diablo 3}.
\newblock Game [PC].   (15 May 2012).
\newblock
\newblock
\shownote{Blizzard Entertainment, Irvine, California, United States.}


\bibitem{GameGallery}
{{Cloudhead Games Ltd.}} 2016.
\newblock \emph{The Gallery - Episode 1: Call of the Starseed}.
\newblock Game [SteamVR].   (5 April 2016).
\newblock
\newblock
\shownote{Cloudhead Games Ltd., Vancouver Island, Canada.}


\bibitem{cmentowski2019outstanding}
{Sebastian Cmentowski}, {Andrey Krekhov}, {and} {Jens Kr{\"u}ger}. 2019.
\newblock \showarticletitle{Outstanding: A Perspective-Switching Technique for
  Covering Large Distances in VR Games}. In {\em Extended Abstracts of the 2019
  CHI Conference on Human Factors in Computing Systems}. ACM, ACM, New York,
  NY, USA, LBW1612.
\newblock


\bibitem{GameGreenHell}
{{Creepy Jar}}. 2018.
\newblock \emph{Green Hell}.
\newblock Game [PC].   (29 August 2018).
\newblock
\newblock
\shownote{Creepy Jar, Warsaw, Poland.}


\bibitem{GameHamurabi}
{Doug Dyment}. 1968.
\newblock \emph{Hamurabi}.
\newblock Game [PC].   (1968).
\newblock


\bibitem{GameFortnite}
{{Epic Games}} {and} {{People Can Fly}}. 2017.
\newblock \emph{Fortnite}.
\newblock Game [PC].   (25 July 2017).
\newblock
\newblock
\shownote{Epic Games and Gearbox Publishing, Raleigh, North Carolina, United
  States.}


\bibitem{hamari2010game}
{Juho Hamari} {and} {Vili Lehdonvirta}. 2010.
\newblock \showarticletitle{Game design as marketing: How game mechanics create
  demand for virtual goods}.
\newblock {\em International Journal of Business Science \& Applied
  Management\/} {5}, 1 (2010), 14--29.
\newblock


\bibitem{krekhov2018gullivr}
{Andrey Krekhov}, {Sebastian Cmentowski}, {Katharina Emmerich}, {Maic Masuch},
  {and} {Jens Kr{\"u}ger}. 2018.
\newblock \showarticletitle{GulliVR: A walking-oriented technique for
  navigation in virtual reality games based on virtual body resizing}. In {\em
  Proceedings of the 2018 Annual Symposium on Computer-Human Interaction in
  Play}. ACM, ACM, New York, NY, USA, 243--256.
\newblock


\bibitem{krekhov2017self}
{Andrey Krekhov}, {Katharina Emmerich}, {Philipp Bergmann}, {Sebastian
  Cmentowski}, {and} {Jens Kr{\"u}ger}. 2017.
\newblock \showarticletitle{Self-transforming controllers for virtual reality
  first person shooters}. In {\em Proceedings of the Annual Symposium on
  Computer-Human Interaction in Play}. ACM, ACM, New York, NY, USA, 517--529.
\newblock


\bibitem{mine1997moving}
{Mark~R Mine}, {Frederick~P Brooks~Jr}, {and} {Carlo~H Sequin}. 1997.
\newblock \showarticletitle{Moving objects in space: exploiting proprioception
  in virtual-environment interaction.}. In {\em SIGGRAPH}, Vol.~97. ACM
  Press/Addison-Wesley Publishing Co., New York, NY, USA, 19--26.
\newblock


\bibitem{GameMinecraft}
{{Mojang}}. 2009.
\newblock \emph{Minecraft}.
\newblock Game [PC].   (17 May 2009).
\newblock
\newblock
\shownote{Mojang, Stockholm, Sweden.}


\bibitem{GameOregonTrail}
{Don Rawitsch}, {Bill Heinemann}, {and} {Paul Dillenberger}. 1971.
\newblock \emph{The Oregon Trail}.
\newblock Game [PC].   (3 December 1971).
\newblock
\newblock
\shownote{Minnesota Educational Computing Consortium and The Learning Company,
  Minnesota.}


\bibitem{GameArkhamVR}
{{Rocksteady Studios}}. 2016.
\newblock \emph{Batman: Arkham VR}.
\newblock Game [PS4 VR].   (13 October 2016).
\newblock
\newblock
\shownote{Warner Bros. Interactive Entertainment, Burbank, California, United
  States.}


\bibitem{Sztajer2010Inventory}
{Paul Sztajer}. 2010.
\newblock Mechanical Breakdown: The Inventory.
\newblock   (August 2010).
\newblock
\showURL{%
Retrieved July 4, 2019 from
  \url{https://kotaku.com/mechanical-breakdown-the-inventory-5612149}}


\bibitem{steamVR}
{{Valve}}. 2014.
\newblock \emph{SteamVR}.
\newblock Software [PC].   (January 2014).
\newblock
\newblock
\shownote{Valve, Bellevue, Washington State, United States.}


\bibitem{wegner2017comparison}
{Konstantin Wegner}, {Sven Seele}, {Helmut Buhler}, {Sebastian Misztal},
  {Rainer Herpers}, {and} {Jonas Schild}. 2017.
\newblock \showarticletitle{Comparison of two inventory design concepts in a
  collaborative virtual reality serious game}. In {\em Extended Abstracts
  Publication of the Annual Symposium on Computer-Human Interaction in Play}.
  ACM, ACM, New York, NY, USA, 323--329.
\newblock


\bibitem{zenner2019drag}
{Andr{\'e} Zenner} {and} {Antonio Kr{\"u}ger}. 2019.
\newblock \showarticletitle{Drag: on: A Virtual Reality Controller Providing
  Haptic Feedback Based on Drag and Weight Shift}. In {\em Proceedings of the
  2019 CHI Conference on Human Factors in Computing Systems}. ACM, ACM, New
  York, NY, USA, 211.
\newblock


\end{thebibliography}

\end{document}